\documentclass[aps,pra,twocolumn]{revtex4-1} 

\usepackage{verbatim}
\usepackage{dsfont} 
\usepackage{amsmath}
\usepackage{amsthm} 
\usepackage{enumerate}
\usepackage{graphicx}
\usepackage{mathtools}
\usepackage{bm} 
\usepackage{amssymb} 
\usepackage{blindtext}
\usepackage{tcolorbox}
\usepackage{appendix} 
\usepackage{hyperref} 
\usepackage[capitalize,nameinlink]{cleveref}
\usepackage{xcolor}
\hypersetup{
    colorlinks=true,
    linkcolor=teal,
    citecolor=magenta,
    unicode=true,  
}
\usepackage{textpos}

\newcommand{\braket}[1]{\left\langle #1 \right\rangle}

\newcommand{\PF}{P_{\mathrm{F}}}
\newcommand{\PR}{P_{\mathrm{R}}}

\usepackage{calc} 
\usepackage{accents}

\usepackage[normalem]{ulem}

\begin{document}

\title{Maxwell's Demon walks into Wall Street:\\Stochastic Thermodynamics meets Expected Utility Theory}

\author{Andr\'es F. Ducuara$^{1,2}$} 
\email[]{andres.ducuara@yukawa.kyoto-u.ac.jp}

\author{Paul Skrzypczyk$^{3,4}$}
\email[]{paul.skrzypczyk@bristol.ac.uk}

\author{Francesco Buscemi$^{5}$}
\email[]{buscemi@nagoya-u.jp}

\author{Peter Sidajaya$^{6}$}
\email[]{peter.sidajaya@u.nus.edu}

\author{Valerio Scarani$^{6,7}$}
\email[]{physv@nus.edu.sg}

\affiliation{$^{1}$Yukawa Institute for Theoretical Physics, Kyoto University, Kitashirakawa Oiwakecho, Sakyo-ku, Kyoto 606-8502, Japan
\looseness=-1}

\affiliation{$^{2}$Center for Gravitational Physics and Quantum Information, Yukawa Institute for Theoretical Physics, Kyoto University
\looseness=-1} 

\affiliation{$^{3}$H.H. Wills Physics Laboratory, University of Bristol, Tyndall Avenue, Bristol, BS8 1TL, United Kingdom 
\looseness=-1}

\affiliation{$^{4}$CIFAR Azrieli Global Scholars program, CIFAR, Toronto, Canada\looseness=-1}

\affiliation{$^{5}$Graduate School of Informatics, Nagoya University, Chikusa-Ku, Nagoya 464-8601, Japan
\looseness=-1}

\affiliation{$^{6}$Centre for Quantum Technologies, National University of Singapore, 3 Science Drive 2, Singapore 117543,
Singapore
\looseness=-1}

\affiliation{$^{7}$Department of Physics, National University of Singapore, 2 Science Drive 3, Singapore 117542,
Singapore
\looseness=-1}

\date{\today}

\begin{abstract}
    The interplay between thermodynamics and information theory has a long history, but its quantitative manifestations are still being explored. We import tools from expected utility theory from economics into stochastic thermodynamics. We prove that, in a process obeying Crooks' fluctuation relations, every $\alpha$ R\'enyi divergence between the forward process and its reverse has the operational meaning of the ``certainty equivalent'' of dissipated work (or, more generally, of entropy production) for a player with risk aversion $r=\alpha-1$. The two known cases $\alpha=1$ and $\alpha=\infty$ are recovered and receive the new interpretation of being associated to a risk-neutral and an extreme risk-averse player respectively. Among the new results, the condition for $\alpha=0$ describes the behavior of a risk-seeking player willing to bet on the transient violations of the second law. Our approach further leads to a generalized Jarzynski equality, and generalizes to a broader class of statistical divergences.
\end{abstract} 
\maketitle

\begin{textblock*}{3cm}(17cm,-11cm)
  \footnotesize YITP-23-71
\end{textblock*}
\vspace{-0.3cm}

\textit{Introduction.---} In the top-down approach of standard thermodynamics, the second law dictates an intrinsic irreversibility in nature. By contrast, dynamics are reversible in the bottom-up approach of statistical mechanics. Stochastic thermodynamics establishes a link between these two approaches: transient violations of the second law are allowed, while the law still holds on average \cite{ST_review_evans, ST_review_seifert1, ST_review_searles, ST_review_Esposito09, ST_review_jarzynski, ST_review_seifert2,  ST_review_vandenbroeck, ST_review_garner, ST_review_peliti, ST_review_paternostro}. Transient violations of the second law were first considered in the works of Evans and Searles \cite{evans1, evans2} in the early 90's. Major breakthroughs were obtained by looking at systems evolving under a Hamiltonian while coupled to a thermal bath at temperature $T$. For this dynamics, the second law takes the form of the Clausius inequality
\begin{equation}\label{e:clausius}
    \braket{W_{\textrm{diss}}}\equiv\braket{W}-\Delta F\,\geq 0,
\end{equation}
stating that, to effect the transition between an initial thermal state with free energy $F_i$ and a final thermal state with free energy $F_f$, one has to invest (on average) an amount of work not smaller than $\Delta F=F_f-F_i$. The extra work that must be provided in a non-quasistatic process is called the \textit{dissipated work}. In 1997, Jarzynski derived a new fluctuation relation, which remarkably is an equality \cite{JE}:
\begin{equation}\label{e:jarzynski}
    \braket{e^{-\beta W_{\textrm{diss}}}}=1,
\end{equation} with $\beta=1/k_BT$ the inverse temperature. Clausius' inequality \eqref{e:clausius} follows from Jarzynski's equality \eqref{e:jarzynski}, by invoking Jensen's inequality $\braket{e^{-x}}\geq e^{-\braket{x}}$. Jarzynski's equality implies that any stochastic violations of the second law are exponentially suppressed. 

Shortly after Jarzynski, Crooks provided an even more refined fluctuation theorem \cite{CFT}:
\begin{equation}\label{eq:crooks}
    \beta W_{\textrm{diss}}\equiv\,w\,=\,\ln\frac{\PF(w)}{\PR(-w)}
\end{equation}
where $\PF(w)$ is the probability density of dissipating an amount of work $w$ (in dimensionless units) in the so-called `forward' (physical) process, and $\PR(-w)$ is the probability density of dissipating an equal and opposite amount of work $-w$ in the so-called `reverse' (or backward) process \footnote{Note that in much of the literature, $\PF(w)$ is instead taken to define the \emph{work done} instead of the dissipated work (and similarly for $\PR(-w)$). Since the dissipated work and work done differ only by the subtraction of the equilibrium free-energy difference $\Delta F$, all results can be stated either in terms of the probability distribution dissipated work or the work done (as is well known in the community). Here, for simplicity of presentation, we will restrict our presentation to dissipated work, without any loss of generality. \textcolor{black}{For the same reasons and for simplicity, we write this dependence using the notation $w \equiv \beta W_{\textrm{diss}}$.}}. The latter is defined by running the Hamiltonian evolution of the system backwards\textcolor{black}{, through a backward trajectory following a time-reversed dynamics}, while interacting with the same thermal bath as in the forward process. \textcolor{black}{Crooks' relation assumes local detailed balance (microreversibility) and, in particular,
can be used to derive Jarzynski's equality  by rewriting} $e^{-\beta w}\PF(w)=\PR(-w)$ and integrating over $w$. Crooks' relation more directly shows that any stochastic violation of the second law ($w < 0$) implies that the associated reverse process necessarily has non-negligible fluctuations that dissipates work.

The above results in stochastic thermodynamics were later connected to information-theoretic quantities. From this perspective, the \emph{average dissipated work} in the physical process
\begin{align}
\beta\braket{W_{\textrm{diss}}}&=\braket{\ln(\PF(w)/\PR(-w))}_{\PF},\nonumber\\
    &=\int dw \PF(w)\ln\frac{\PF(w)}{\PR(-w)},\nonumber \\&\equiv D(\PF(w)\|\PR(-w)),\label{eq:WKL}
\end{align} is immediately recognized as the Kullback-Leibler (KL) divergence (relative entropy), as first noticed by Kawai et al.~\cite{KPB1}. In \eqref{eq:WKL}, and in what follows, when there is any ambiguity, we will denote which probability distribution the average is taken over by using a final subscript. More recently, Yunger-Halpern et al.~proved that, when \eqref{eq:crooks} holds, the \emph{worst-case dissipated work} (largest fluctuation) is also measured by a statistical divergence between the forward and the backward statistics \cite{NYH1}:
\begin{eqnarray}
    \beta W_{\textrm{diss}}^{\textrm{worst}}&=&\ln\left(\min\{\lambda:\,\PF(w)\leq \lambda \PR(-w)\;\forall w\}\right),\nonumber\\ &\equiv&D_\infty(\PF(w)\|\PR(-w))\,,\label{eq:Nicole}
\end{eqnarray}
which is also sometimes referred to as the \emph{max relative entropy}. Both of these divergences belong to the family of R\'enyi divergences \cite{NV1, CT}
\begin{equation}
    D_\alpha(P\|Q)=\frac{1}{\alpha-1}\ln\left\langle\left(\frac{P}{Q}\right)^{\alpha-1}\right\rangle_{P}\,,
\end{equation} that are proper divergences for $\alpha\geq 0$ (but can be mathematically defined for all values of $\alpha$), the KL divergence being the special case $\alpha=1$, and $D_\infty(P\|Q)$ (as in \eqref{eq:Nicole}) obtained by taking the limit $\alpha \to \infty$. 

This raises an intriguing question: whether the R\'enyi divergences have any operational significance beyond the cases $\alpha = 1$ and $\alpha = \infty$ from above. In this paper, we show that this is indeed the case: \textit{every R\'enyi divergence comparing the forward and reverse statistics of dissipated work has an operational meaning}. Our result  reads
\begin{equation}\label{eq:main}
    \beta W_{\textrm{diss},r}^{\textrm{CE}}=D_{1+r}(\PF(w)\|\PR(-w)),
\end{equation} where CE stands for ``\textit{certainty equivalent}'' and $r$ is a parameter that quantifies the \textit{risk aversion} of an agent. These are standard notions from \textit{expected utility theory} (EUT) -- widely used in the economic sciences. Since most physicists may not be familiar with this, we provide next a brief introduction to these ideas, before presenting our results in detail.

\textit{Basics of EUT.---} The theory of expected utility is the study of rational agents acquiring, trading, and hoarding assets such as wealth, goods, and services. First formalised by von Neumann and Morgenstern within the theory of games and economic behaviour back in 1944 \cite{risk_vNM}, it witnessed major developments during the 50's and 60's, so as to include decision-making rational agents choosing between alternatives involving uncertainty, as well as descriptions of behavioural tendencies such as risk-aversion \cite{risk_bernoulli, risk_arrow, risk_pratt, risk_finetti}, and connections to information theory \cite{kelly}. Recently, further connections between expected utility theory, information theory, and quantum resource theories have been put forward \cite{DS2022, risk_soklakov, BLP1}. Since dissipated work is a type of liability rather than an asset, a good analogy is that of a \textit{tax game}. A referee (tax collector) gives the player (contributor) two options to choose between: (i) The first alternative corresponds to tossing a fair coin, with heads meaning the player pays $\$1000$ in tax, and tails meaning the player obtains a tax rebate of $\$100$ (i.e.~pays $-\$100$ in tax); (ii) The second alternative corresponds to the player paying a \textit{fixed amount} of tax (between $-\$100$ and $\$1000$, of course). The expected loss of the first -- stochastic  -- choice  is $\$450$, so a player who would choose option 2 even when the fixed amount of tax is \emph{larger} than this average should naturally be considered a \emph{risk-averse} player; they are willing to pay out above the average in order to avoid the possibility of the large tax bill. On the other hand, a player offered to pay a fixed  amount of tax \emph{smaller} than this average, and still choosing the first option to gamble, should be considered a \emph{risk-seeking} player. The so-called \textit{certainty equivalent} (CE) is the fixed amount offered in option 2, for which the player considers the two alternatives equivalent. The value of the certainty equivalent tax can thus be seen to quantify the risk behaviour of the agent.

\textcolor{black}{To make the analysis more concrete, a rational agent is modelled by a \textit{utility function}, which represents the level of satisfaction they receive from some alternative $t$ by $u(t)$. In this work we consider decision problems where the rational agent is asked to pay the referee in terms of energy, and so the utility function must naturally be \emph{decreasing} (the more there is to be paid the less satisfied the agent is going to be). Utility functions are assumed be twice differentiable, modelling the fact that smooth changes in the alternative should intuitively produce smooth changes in the agent's satisfaction. A highly nontrivial conceptual breakthrough developed by economists during the 50's and 60's established that an agent's aversion to risk is encoded in the \emph{curvature} of their utility function as \cite{risk_arrow, risk_pratt, risk_finetti, DS2022}:}
\begin{equation}
    \begin{array}{ll}
    \textrm{Risk-averse:}&\textrm{$u$ concave i.e. } u{''}(t)<0\\
    \textrm{Risk-neutral:}&\textrm{$u$ linear i.e. } u{''}(t)=0\\
    \textrm{Risk-seeking:}&\textrm{$u$ convex i.e. } u{''}(t)>0\,.
    \end{array}
\end{equation}
One of the basic measures of risk aversion is the \emph{Absolute Risk Aversion} (ARA), given by \footnote{Note that it is more standard to see the ARA defined with an additional minus sign. Here, we since we are considering liabilities instead of gains, it is natural to omit the minus sign. This means that risk-averse agents will have a positive ARA, while risk-seeking agents will have a negative ARA, in line with the definitions for gains. }
\begin{align}
    \textrm{ARA}_u(t)
    \coloneqq
    \frac{
    u{''}(t)
    }{
    u{'}(t)
    }
    .
\end{align}
In view of the fact that below we will apply EUT to the case of dissipated work $w$, which is a liability like tax, we will focus on strictly decreasing utility functions ($u{'}(t)<0$). 
Thus, we  have $\textrm{ARA}>0$ for risk-averse agents, $\textrm{ARA}=0$ for risk-neutral agents and $\textrm{ARA}<0$ for risk-seeking agents. Finally, given a utility function, the certainty equivalent tax $T^{\rm CE}$ is defined implicitly via
\begin{equation}\label{eq:CE}
    u(T^{\rm CE})\coloneqq \langle u(t)\rangle\,,
\end{equation}
i.e.~the amount of tax that has the same utility as the average utility, and is therefore equally preferable to the stochastic situation.

For our purposes, we will focus on utility functions $u_r$ with constant ARA (CARA), meaning that $\textrm{ARA}_{u_r}(t)=r$ for all $t$. This widely-studied and particularly simple utility function models agents whose risk tendency is constant, independent of the size of their liability. By solving the differential equation, one finds that this family of utility functions is given by 
\begin{align}
    u_r(t)
    =
    \begin{cases}
        \frac{1}{r}
        \left(
        1-
        e^{rt}
        \right)
        , 
        & \text{if}\ r \neq 0 \\
        -t, 
        & \text{if}\ r = 0.
    \end{cases}
    \label{eq:CARA}
\end{align} We note that utility functions have built in a type of gauge invariance, whereby two utility functions which differ only by scale and shift represent the same agent (risk tendency). Here we have used this freedom to fix the function at zero liability: $u_r(0)=0$ and $u'_r(0)=-1$.

\begin{figure}[t!]
    \centering
    \includegraphics[scale=0.94]{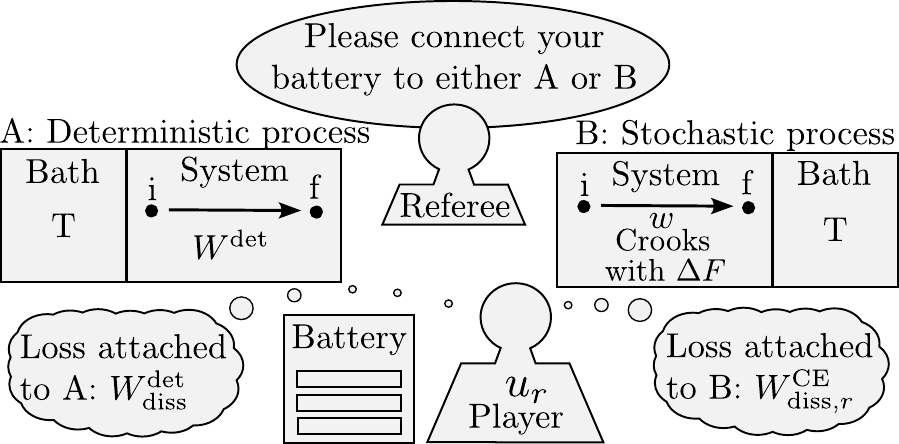}
    \caption{Representation of the work dissipation game. The player, modelled by a CARA utility function $u_r$ \eqref{eq:CARA}, is asked to use their battery to implement either (i) a deterministic process (which will require a fixed amount of work $W^{\textrm{det}}$), or (ii) a stochastic process in which work follows Crooks' fluctuation relation \eqref{eq:crooks}with free energy difference $\Delta F$. Risk-averse players ($r>0$) will choose the deterministic process unless $W^{\textrm{det}}$ is too large; risk-seeking players ($r<0$) will choose the stochastic process unless $W^{\textrm{det}}$ is sufficiently low. The certainty equivalent determines the value at which the player switches: for $W^{\textrm{det}}-\Delta F\equiv W_{\rm diss}^{\textrm{det}}>W_{\textrm{diss},r}^{\textrm{CE}}$, the player chooses the stochastic process; if the inequality is in the opposite direction, they choose the deterministic one.}
    \label{fig:dp1}
\end{figure}

\textit{EUT and dissipated work.---} With the above basics of EUT in place, we are now in a position to make the connection with stochastic thermodynamics. The tax game described above can be generalised and re-narrated as a \emph{work dissipation game} (\cref{fig:dp1}). 

The player is in possession of charged battery, which they can use in order to carry out a physical process. The referee gives the player two options to choose between: (i) The player carries out a physical process which will dissipate a stochastic amount of work $\beta W_{\textrm{diss}}\equiv w$ according to $\PF(w)$, and obeying Crooks' fluctuation relation \eqref{eq:crooks} with equilibrium free energy difference $\Delta F$; (ii) The player carries out an alternative physical process which dissipates a \emph{deterministic} amount of work $W_{\rm diss}^{\textrm{det}}$.

Which of the two options is more preferable to a player will depend upon their utility function $u_r(w)$, and in particular the \emph{certainty equivalent dissipated work} $W_{\textrm{diss},r}^{\textrm{CE}}$ it defines: the player will choose the deterministic process if the amount of work it deterministically dissipates is smaller than the certainty equivalent, $W_{\rm diss}^{\textrm{det}}~<~W_{\textrm{diss},r}^{\textrm{CE}}$; on the other hand, they will choose the stochastic process if $W_{\rm diss}^{\textrm{det}}>W_{\textrm{diss},r}^{\textrm{CE}}$ (and of course the two are equally preferable if the certainty equivalent exactly coincides with $W_{\rm diss}^{\textrm{det}}$). Applying \eqref{eq:CE} to our context, we have
\begin{align}\label{e:CE diss work}
    u_r
    \left(
    \beta W_{\textrm{diss},r}^{\textrm{CE}}
    \right)
    =
    \left\langle u_r(w)\right\rangle_{\PF}\,.
\end{align} 
Therefore
\begin{align}
    \beta W_{{\rm diss},r}^{\textrm{CE}}&= u^{-1}_{r}\left(\left\langle u_r(w)\right\rangle_{\PF}\right),\nonumber
    \\
    &=u^{-1}_{r}
    \left(
        \frac{1}{r}
        \left\langle
            1-\left( \frac{\PF(w)}{\PR(-w)}\right)^{r}
        \right\rangle_{P_F}
    \right),\label{interm1}
    \\
    &=u^{-1}_{r}
    \left(
        \frac{1}{r}
        \left[1-\int dw
            \PF(w)\left(\frac{\PF(w)}{\PR(-w)}\right)^{r}
            \right]
    \right),\nonumber\\
    &=\frac{1}{r}
    \ln
    \left[
        \int dw
            \PF(w)
            \left(\frac{\PF(w)}{\PR(-w)}\right)^{r}
    \right],\label{interm2}
    \\
    &=
    D_{1+r}\left(\PF(w)\|\PR(-w)\right),\label{e:final result}
\end{align}
which is the announced main result \eqref{eq:main}. To reach \eqref{interm1}, we inserted \eqref{eq:crooks} and the definition \eqref{eq:CARA} of $u_r$, while for \eqref{interm2} we used that the inverse of $u_r$ is
\begin{align}
    u_r^{-1}(x)
    =
    \begin{cases}
        \frac{1}{r}
        \ln\left(
        1-rx
        \right)
        , 
        & \text{if}\ r \neq 0, \\
        -x, 
        & \text{if}\ r = 0.
    \end{cases}\,
    \label{eq:CARAinv}
\end{align}

To analyse this result, we keep in mind that 
$D_{\alpha}\leq D_{\beta}$ for $\alpha\leq \beta$, and that $D_0(P_1\|P_2)=0$ (if $P_1$ and $P_2$ have the same support; here, we routinely consider distributions with full support). As desired, $W_{{\rm diss},r}^{\rm{CE}}$ will be higher (lower) the more risk-averse (risk-seeking) the player is. Let us now look at the different player behaviors.  

\textit{Risk-neutral players} will switch their choice by comparing $W^{\textrm{det}}_{\rm diss}$ with the average dissipated work $\langle W_{\rm diss}\rangle$, since in this case the certainty equivalent $W_{{\rm diss},r=0}^{\rm{CE}}~=~\braket{W_{\rm diss}}$ [Eq.~\eqref{eq:WKL}]. This is as expected. 

\textit{Risk-averse players} by default like the deterministic option, but will switch to the stochastic one if $W^{\textrm{det}}_{\rm diss}\geq~ W_{{\rm diss},r>0}^{\rm{CE}}>\braket{W_{\rm diss}}$. A case worth a special mention is the limit $r\rightarrow\infty$, where Eq.~(\ref{eq:Nicole}) is recovered \cite{NYH1}. We can thus give a new meaning to this result: in this limit, players are so risk-averse that they would only switch to the stochastic option if the deterministic process offered to them would dissipate with certainty more work than \textit{any} fluctuation (including the worst case fluctuation) does. 

\textit{Risk-seeking players} by default would enjoy the stochastic option, but will switch to the deterministic one if $W^{\textrm{det}}_{\rm diss}\leq W_{{\rm diss},r<0}^{\rm{CE}}<\braket{W_{\rm diss}}$. The zoology of risk-seekers is richer. Players with $-1<r<0$ change their choice for positive, but less than average, dissipated work. Players with $r=-1$ switch choice precisely at $W^{\textrm{det}}_{\rm diss}=0$: these are the players willing to bet on the transient violations of the second law. Finally, our derivation still holds formally even for when $r<-1$: while the r.h.s.~of \eqref{eq:main} can no longer be interpreted as a divergence since $\alpha=1+r<0$, one can prove that $D_\alpha$ becomes more and more negative with $\alpha$ \cite{NV1}. Again, this describes what we expect: players so extreme, that they change their choice at negative dissipated work, i.e.~they bet on large transient violations of the second law. In the limit, as $r \to -\infty$, we find a counterpart to Eq.~\eqref{eq:Nicole}: the largest transient violation of the second law $W_{\rm diss}^{\rm viol}$ can be shown to be given by 
\begin{eqnarray}
    \beta W_{\textrm{diss}}^{\textrm{viol}}&=&\ln\left(\max\{\lambda:\,\lambda\PR(-w)\leq \PF(w)\;\forall w\}\right),\nonumber\\ &\equiv&D_{-\infty}(\PF(w)\|\PR(-w))\,.\label{eq:viol}
\end{eqnarray}
In our setting, we can also give an alternative meaning to this result: in this limit, players are so risk-seeking that they would only switch to the deterministic option if it would, with certainty, give them a violation of the second law larger than any fluctuation of the stochastic process. 

\textit{Generalized Jarzynski equality.}---It is also possible to use our results to rederive, and shed light on, a generalized form of Jarzynski equality. From \eqref{eq:CARA}, after simply rearrangement it follows that $\langle e^{r\beta W_{\rm diss}}\rangle_{\PF} = (1-r\langle u_r(w)\rangle_{\PF})$. Combining this with \eqref{e:CE diss work} and \eqref{e:final result}, we obtain
\begin{align}
    \langle e^{r\beta W_{\rm diss}}\rangle_{\PF} &= e^{r \beta W_{{\rm diss},r}^{\textrm{CE}}},\label{e:J1}\\
    &= e^{r D_{1+r}\left(\PF(w)\|\PR(-w)\right)}.\label{e:J2}
\end{align}
The second equality \eqref{e:J2} recovers the main result of \cite{Wei_Plenio} (although written in a slightly different form), which we now see holds assuming only Crook's fluctuation relation \eqref{eq:crooks}. The first equality is novel, and shows that we can give the r.h.s. of this generalized Jarzynski equality an operational interpretation, in terms of the certainty equivalent dissipated work. The equality \eqref{eq:WKL} is recovered for $r=0$ (after Taylor expansion), while the original Jarzynski equality \eqref{e:jarzynski} is recovered for $r=-1$ because $W_{{\rm diss},r=-1}^{\textrm{CE}} = 0$. The latter is also the only case, in which the r.h.s.~is independent of the physical process: for any other value of $r$, the r.h.s.~depends upon the process, solely through the certainty equivalent dissipated work. 

\textit{Generalisation to entropy production.---} So far we have presented our results in the most well studied scenario in stochastic thermodynamics: that of a system evolving under a Hamiltonian while coupled to a thermal bath, for which the second law becomes Clausius' inequality, and in which the original theorems of Jarzynski \cite{JE} and Crooks \cite{CFT} were formulated. However, stochastic thermodynamics has gone beyond that specific physical scenario. Consider a stochastic two-time process, described by the probability distribution $\PF(\lambda_i,\lambda_f)$ of starting in state $\lambda_i$ and finishing in state $\lambda_f$. For any such process, \textit{detailed entropy production} is defined as \cite{ST_review_evans, ST_review_searles, ST_review_Esposito09, ST_review_seifert2, ST_review_paternostro}
\begin{equation}\label{eq:crooksgen}
    s(\lambda_i,\lambda_f)\,=\,\ln\frac{\PF(\lambda_i,\lambda_f)}{\PR(\lambda_f,\lambda_i)},
\end{equation} 
where the reverse process $\PR$ can be defined through a generic logical recipe of retrodiction \cite{BS2021,ABS}. As defined, $s$ is dimensionless; the thermodynamic entropy is $S=\beta^{-1} s$; and indeed, for the scenario of Jarzynski and Crooks, $\beta W_{\rm diss}=S$ holds, so that \eqref{eq:crooksgen} becomes \eqref{eq:crooks}. This generic form of entropy production obeys the second law on average since
\begin{eqnarray}
\braket{s}&=&D(\PF(\lambda_i,\lambda_f)\|\PR(\lambda_f,\lambda_i))\label{eq:SKL}\geq 0\,,
\end{eqnarray} but some transitions $\lambda_i\rightarrow \lambda_f$ are such that $s(\lambda_i,\lambda_f)<0$. It also obeys a generalized Jarzynski equality $\braket{e^{-s}}=1$. 

Our approach based upon EUT carries over unmodified to this very general setting. The generalization of \eqref{eq:main} states that \textit{the certainty equivalent entropy production of any stochastic process} satisfying detailed entropy production \eqref{eq:crooksgen} for an agent with constant absolute risk aversion $r$ satisfies
\begin{equation}
s^{\textrm{CE}}_r=D_{1+r}\left(\PF(\lambda_i,\lambda_f)\|\PR(\lambda_f,\lambda_i)\right),
\end{equation}
where $s^{\textrm{CE}}_r= u^{-1}_{r}\left(\left\langle u_r(s)\right\rangle_{\PF}\right)$.
The previous analysis also carries over: so, for instance, players with $r=-1$ are risk-seeking to the point of betting on the transient violations of the second law. The fully generalized Jarzynski equality similarly becomes
\begin{equation}
    \langle e^{rs}\rangle = e^{r s_{r}^{\textrm{CE}}}= e^{r D_{1+r}\left(\PF(\lambda_i,\lambda_f)\|\PR(\lambda_f,\lambda_i)\right)},\label{e:Jg}
\end{equation}
which again has the feature that only for $r=-1$ does the r.h.s. become independent of the physical process taking place. 

\textit{Generalization to $f$-divergences.}--- It is also possible to extend the results above beyond R\'enyi divergences, to more general $f$-divergences \cite{FD}. We saw above that if we model agents using the CARA utility function \eqref{eq:CARA}, then the certainty equivalent dissipated work (or entropy production) is given by a R\'enyi divergence. If instead we consider a more general class of utility functions of the form
\begin{equation}
    v_f(s) = f(e^{-s}),
\end{equation}
where $f$ is convex and satisfies $f(1) = 0$, then it follows that
\begin{align}
    s^{\textrm{CE}}_f &= v_f^{-1}\left(\langle v_f(s)\rangle_{\PF}\right),\nonumber \\
    &= v_f^{-1}\left(\int ds \PF(\lambda_i,\lambda_f) f\left(\frac{\PR(\lambda_f,\lambda_i)}{\PF(\lambda_i,\lambda_f)}\right)\right),\nonumber \\
    &= v_f^{-1}\left(\int ds \PR(\lambda_f,\lambda_i) \hat{f}\left(\frac{\PF(\lambda_i,\lambda_f)}{\PR(\lambda_f,\lambda_i)}\right)\right),\nonumber \\
    &= v_f^{-1}\left(
    \mathcal{D}_{\hat{f}}
    (
    \PF(\lambda_i,\lambda_f)\| \PR(\lambda_f,\lambda_i)
    )
    \right) \nonumber \\
    &= - \ln 
    \left( f^{-1}
    \left(
    \mathcal{D}_{\hat{f}}
    (
    \PF(\lambda_i,\lambda_f)\| \PR(\lambda_f,\lambda_i) 
    )
    \right) \right)
\end{align}
where
\begin{equation}
    \mathcal{D}_f(P||Q):=\int dx Q(x) f\left(\frac{P(x)}{Q(x)}\right) 
\end{equation}
is the $f$-divergence and $\hat{f}(x):=xf(x^{-1})$ is the perspective function of $f$, and is defined as such so that $D_f(P||Q) = D_{\hat{f}}(Q||P)$. Our main result (\ref{e:final result}) is then recovered with $f(x)=\frac{1}{r}(1-x^{-r})$.

\textit{Conclusion.---} In this paper, we provided an operational meaning for all the $\alpha$-R\'enyi divergences between the statistics of a stochastic thermodynamical process and those of its reverse. Contrary to what one might have expected from the basic relation \eqref{eq:WKL}, the R\'enyi divergences does not arise as the expectation value of a suitable function $f_\alpha$ of the thermodynamical variable (dissipated work, entropy production). Rather, it is the certainty equivalent of that variable for an agent, whose behavior is captured by a utility function with constant absolute risk aversion $r=\alpha-1$. This connection between information theory and stochastic thermodynamics was made possible by importing the theoretical machinery of the economic theory of rational agents. \textcolor{black}{The certainty equivalent is the quantity that provides resolution to decision problems, as it determines the alternative that agents would choose when facing decision problems involving uncertainty. In this latter sense, the certainty equivalent acquires an operational meaning which can be measured in experimental setups involving decision-making agents. The relationship we derived connects the fields of stochastic thermodynamics, information theory, the theory of expected utility theory and, as such, can naturally be integrated and further exploited, for instance, within the field of information thermodynamics \cite{sagawa2013}. Similarly, whilst we have considered here a process between two points in phase space (as in a single forward or backward process), one can also imagine considering a multi-step processes, such as thermal cycles, and similarly explore the applicability of these ideas for such thermal machines.}

\textcolor{black}{Recently, some works have also addressed thermodynamics using economic-theoretic language \cite{morales23, caravelli22}. It would be interesting to analyse whether these results can collectively be analysed within a single economic-theoretic framework. Finally, it is also interesting to explore potential applications of utility theory to areas like coding theory, where information-theoretic quantities also emerge \cite{campbell65}.} By walking into Wall Street, if not wealthier, Maxwell's Demon has become more aware of the rationale of its behavior. It may soon invite its quantum alter-ego to take the tour.

\emph{Acknowledgements.} P. Sk. thanks Mate Arcos and Jonathan Oppenheim for
interesting discussions, especially about their independent unpublished work connecting gambling and thermodynamics. A.F.D.~acknowledges support from the International Research Unit of Quantum Information, Kyoto University, the Center for Gravitational Physics and Quantum Information (CGPQI), and from COLCIENCIAS 756-2016. P.Sk.~acknowledges support from a Royal Society URF (NFQI) and is a CIFAR Azrieli Global Scholar in the Quantum Information Science Programme. F.B.~acknowledges support from the MEXT-JSPS Grant-in-Aid for Transformative Research Areas (A) ``Extreme Universe'', No. 21H05183; from the MEXT Quantum Leap Flagship Program (MEXT QLEAP) Grant No. JPMXS0120319794;
from JSPS KAKENHI, Grants No. 20K03746 and No. 23K03230. P.Si.~and V.S.~are supported by the National Research Foundation, Singapore and A*STAR under its CQT Bridging Grant. 

\bibliographystyle{apsrev4-1}
\bibliography{bibliography.bib}

\begin{thebibliography}{37}%
\makeatletter
\providecommand \@ifxundefined [1]{%
 \@ifx{#1\undefined}
}%
\providecommand \@ifnum [1]{%
 \ifnum #1\expandafter \@firstoftwo
 \else \expandafter \@secondoftwo
 \fi
}%
\providecommand \@ifx [1]{%
 \ifx #1\expandafter \@firstoftwo
 \else \expandafter \@secondoftwo
 \fi
}%
\providecommand \natexlab [1]{#1}%
\providecommand \enquote  [1]{``#1''}%
\providecommand \bibnamefont  [1]{#1}%
\providecommand \bibfnamefont [1]{#1}%
\providecommand \citenamefont [1]{#1}%
\providecommand \href@noop [0]{\@secondoftwo}%
\providecommand \href [0]{\begingroup \@sanitize@url \@href}%
\providecommand \@href[1]{\@@startlink{#1}\@@href}%
\providecommand \@@href[1]{\endgroup#1\@@endlink}%
\providecommand \@sanitize@url [0]{\catcode `\\12\catcode `\$12\catcode
  `\&12\catcode `\#12\catcode `\^12\catcode `\_12\catcode `\%12\relax}%
\providecommand \@@startlink[1]{}%
\providecommand \@@endlink[0]{}%
\providecommand \url  [0]{\begingroup\@sanitize@url \@url }%
\providecommand \@url [1]{\endgroup\@href {#1}{\urlprefix }}%
\providecommand \urlprefix  [0]{URL }%
\providecommand \Eprint [0]{\href }%
\providecommand \doibase [0]{http://dx.doi.org/}%
\providecommand \selectlanguage [0]{\@gobble}%
\providecommand \bibinfo  [0]{\@secondoftwo}%
\providecommand \bibfield  [0]{\@secondoftwo}%
\providecommand \translation [1]{[#1]}%
\providecommand \BibitemOpen [0]{}%
\providecommand \bibitemStop [0]{}%
\providecommand \bibitemNoStop [0]{.\EOS\space}%
\providecommand \EOS [0]{\spacefactor3000\relax}%
\providecommand \BibitemShut  [1]{\csname bibitem#1\endcsname}%
\let\auto@bib@innerbib\@empty
\bibitem [{\citenamefont {Evans}\ and\ \citenamefont
  {Searles}(2002)}]{ST_review_evans}%
  \BibitemOpen
  \bibfield  {author} {\bibinfo {author} {\bibfnamefont {D.~J.}\ \bibnamefont
  {Evans}}\ and\ \bibinfo {author} {\bibfnamefont {D.~J.}\ \bibnamefont
  {Searles}},\ }\href {\doibase 10.1080/00018730210155133} {\bibfield
  {journal} {\bibinfo  {journal} {Advances in Physics}\ }\textbf {\bibinfo
  {volume} {51}},\ \bibinfo {pages} {1529} (\bibinfo {year}
  {2002})}\BibitemShut {NoStop}%
\bibitem [{\citenamefont {Seifert}(2008)}]{ST_review_seifert1}%
  \BibitemOpen
  \bibfield  {author} {\bibinfo {author} {\bibfnamefont {U.}~\bibnamefont
  {Seifert}},\ }\href {\doibase 10.1140/epjb/e2008-00001-9} {\bibfield
  {journal} {\bibinfo  {journal} {The European Physical Journal B}\ }\textbf
  {\bibinfo {volume} {64}},\ \bibinfo {pages} {423} (\bibinfo {year}
  {2008})}\BibitemShut {NoStop}%
\bibitem [{\citenamefont {Sevick}\ \emph {et~al.}(2008)\citenamefont {Sevick},
  \citenamefont {Prabhakar}, \citenamefont {Williams},\ and\ \citenamefont
  {Searles}}]{ST_review_searles}%
  \BibitemOpen
  \bibfield  {author} {\bibinfo {author} {\bibfnamefont {E.}~\bibnamefont
  {Sevick}}, \bibinfo {author} {\bibfnamefont {R.}~\bibnamefont {Prabhakar}},
  \bibinfo {author} {\bibfnamefont {S.~R.}\ \bibnamefont {Williams}}, \ and\
  \bibinfo {author} {\bibfnamefont {D.~J.}\ \bibnamefont {Searles}},\ }\href
  {\doibase 10.1146/annurev.physchem.58.032806.104555} {\bibfield  {journal}
  {\bibinfo  {journal} {Annual Review of Physical Chemistry}\ }\textbf
  {\bibinfo {volume} {59}},\ \bibinfo {pages} {603} (\bibinfo {year}
  {2008})}\BibitemShut {NoStop}%
\bibitem [{\citenamefont {Esposito}\ \emph {et~al.}(2009)\citenamefont
  {Esposito}, \citenamefont {Harbola},\ and\ \citenamefont
  {Mukamel}}]{ST_review_Esposito09}%
  \BibitemOpen
  \bibfield  {author} {\bibinfo {author} {\bibfnamefont {M.}~\bibnamefont
  {Esposito}}, \bibinfo {author} {\bibfnamefont {U.}~\bibnamefont {Harbola}}, \
  and\ \bibinfo {author} {\bibfnamefont {S.}~\bibnamefont {Mukamel}},\ }\href
  {\doibase 10.1103/RevModPhys.81.1665} {\bibfield  {journal} {\bibinfo
  {journal} {Rev. Mod. Phys.}\ }\textbf {\bibinfo {volume} {81}},\ \bibinfo
  {pages} {1665} (\bibinfo {year} {2009})}\BibitemShut {NoStop}%
\bibitem [{\citenamefont {Jarzynski}(2011)}]{ST_review_jarzynski}%
  \BibitemOpen
  \bibfield  {author} {\bibinfo {author} {\bibfnamefont {C.}~\bibnamefont
  {Jarzynski}},\ }\href {\doibase 10.1146/annurev-conmatphys-062910-140506}
  {\bibfield  {journal} {\bibinfo  {journal} {Annual Review of Condensed Matter
  Physics}\ }\textbf {\bibinfo {volume} {2}},\ \bibinfo {pages} {329} (\bibinfo
  {year} {2011})}\BibitemShut {NoStop}%
\bibitem [{\citenamefont {Seifert}(2012)}]{ST_review_seifert2}%
  \BibitemOpen
  \bibfield  {author} {\bibinfo {author} {\bibfnamefont {U.}~\bibnamefont
  {Seifert}},\ }\href {\doibase 10.1088/0034-4885/75/12/126001} {\bibfield
  {journal} {\bibinfo  {journal} {Reports on Progress in Physics}\ }\textbf
  {\bibinfo {volume} {75}},\ \bibinfo {pages} {126001} (\bibinfo {year}
  {2012})}\BibitemShut {NoStop}%
\bibitem [{\citenamefont {den Broeck}\ and\ \citenamefont
  {Esposito}(2015)}]{ST_review_vandenbroeck}%
  \BibitemOpen
  \bibfield  {author} {\bibinfo {author} {\bibfnamefont {C.~V.}\ \bibnamefont
  {den Broeck}}\ and\ \bibinfo {author} {\bibfnamefont {M.}~\bibnamefont
  {Esposito}},\ }\href {\doibase 10.1016/j.physa.2014.04.035} {\bibfield
  {journal} {\bibinfo  {journal} {Physica A: Statistical Mechanics and its
  Applications}\ }\textbf {\bibinfo {volume} {418}},\ \bibinfo {pages} {6}
  (\bibinfo {year} {2015})}\BibitemShut {NoStop}%
\bibitem [{\citenamefont {Garner}(2018)}]{ST_review_garner}%
  \BibitemOpen
  \bibfield  {author} {\bibinfo {author} {\bibfnamefont {A.~J.~P.}\
  \bibnamefont {Garner}},\ }in\ \href {\doibase 10.1007/978-3-319-99046-0_27}
  {\emph {\bibinfo {booktitle} {Fundamental Theories of Physics}}}\ (\bibinfo
  {publisher} {Springer International Publishing},\ \bibinfo {year} {2018})\
  pp.\ \bibinfo {pages} {651--679}\BibitemShut {NoStop}%
\bibitem [{\citenamefont {Peliti}\ and\ \citenamefont
  {Pigolotti}(2021)}]{ST_review_peliti}%
  \BibitemOpen
  \bibfield  {author} {\bibinfo {author} {\bibfnamefont {L.}~\bibnamefont
  {Peliti}}\ and\ \bibinfo {author} {\bibfnamefont {S.}~\bibnamefont
  {Pigolotti}},\ }\href@noop {} {\emph {\bibinfo {title} {Stochastic
  Thermodynamics}}}\ (\bibinfo  {publisher} {Princeton University Press},\
  \bibinfo {address} {Princeton, NJ},\ \bibinfo {year} {2021})\BibitemShut
  {NoStop}%
\bibitem [{\citenamefont {Landi}\ and\ \citenamefont
  {Paternostro}(2021)}]{ST_review_paternostro}%
  \BibitemOpen
  \bibfield  {author} {\bibinfo {author} {\bibfnamefont {G.~T.}\ \bibnamefont
  {Landi}}\ and\ \bibinfo {author} {\bibfnamefont {M.}~\bibnamefont
  {Paternostro}},\ }\href {\doibase 10.1103/RevModPhys.93.035008} {\bibfield
  {journal} {\bibinfo  {journal} {Rev. Mod. Phys.}\ }\textbf {\bibinfo {volume}
  {93}},\ \bibinfo {pages} {035008} (\bibinfo {year} {2021})}\BibitemShut
  {NoStop}%
\bibitem [{\citenamefont {Evans}\ \emph {et~al.}(1993)\citenamefont {Evans},
  \citenamefont {Cohen},\ and\ \citenamefont {Morriss}}]{evans1}%
  \BibitemOpen
  \bibfield  {author} {\bibinfo {author} {\bibfnamefont {D.~J.}\ \bibnamefont
  {Evans}}, \bibinfo {author} {\bibfnamefont {E.~G.~D.}\ \bibnamefont {Cohen}},
  \ and\ \bibinfo {author} {\bibfnamefont {G.~P.}\ \bibnamefont {Morriss}},\
  }\href {\doibase 10.1103/PhysRevLett.71.2401} {\bibfield  {journal} {\bibinfo
   {journal} {Phys. Rev. Lett.}\ }\textbf {\bibinfo {volume} {71}},\ \bibinfo
  {pages} {2401} (\bibinfo {year} {1993})}\BibitemShut {NoStop}%
\bibitem [{\citenamefont {Evans}\ and\ \citenamefont {Searles}(1994)}]{evans2}%
  \BibitemOpen
  \bibfield  {author} {\bibinfo {author} {\bibfnamefont {D.~J.}\ \bibnamefont
  {Evans}}\ and\ \bibinfo {author} {\bibfnamefont {D.~J.}\ \bibnamefont
  {Searles}},\ }\href {\doibase 10.1103/PhysRevE.50.1645} {\bibfield  {journal}
  {\bibinfo  {journal} {Phys. Rev. E}\ }\textbf {\bibinfo {volume} {50}},\
  \bibinfo {pages} {1645} (\bibinfo {year} {1994})}\BibitemShut {NoStop}%
\bibitem [{\citenamefont {Jarzynski}(1997)}]{JE}%
  \BibitemOpen
  \bibfield  {author} {\bibinfo {author} {\bibfnamefont {C.}~\bibnamefont
  {Jarzynski}},\ }\href {\doibase 10.1103/PhysRevLett.78.2690} {\bibfield
  {journal} {\bibinfo  {journal} {Phys. Rev. Lett.}\ }\textbf {\bibinfo
  {volume} {78}},\ \bibinfo {pages} {2690} (\bibinfo {year}
  {1997})}\BibitemShut {NoStop}%
\bibitem [{\citenamefont {Crooks}(1999)}]{CFT}%
  \BibitemOpen
  \bibfield  {author} {\bibinfo {author} {\bibfnamefont {G.~E.}\ \bibnamefont
  {Crooks}},\ }\href {\doibase 10.1103/PhysRevE.60.2721} {\bibfield  {journal}
  {\bibinfo  {journal} {Phys. Rev. E}\ }\textbf {\bibinfo {volume} {60}},\
  \bibinfo {pages} {2721} (\bibinfo {year} {1999})}\BibitemShut {NoStop}%
\bibitem [{Note1()}]{Note1}%
  \BibitemOpen
  \bibinfo {note} {Note that in much of the literature, $P_{\protect \mathrm
  {F}}(w)$ is instead taken to define the \protect \emph {work done} instead of
  the dissipated work (and similarly for $P_{\protect \mathrm {R}}(-w)$). Since
  the dissipated work and work done differ only by the subtraction of the
  equilibrium free-energy difference $\Delta F$, all results can be stated
  either in terms of the probability distribution dissipated work or the work
  done (as is well known in the community). Here, for simplicity of
  presentation, we will restrict our presentation to dissipated work, without
  any loss of generality. \textcolor {black}{For the same reasons and for
  simplicity, we write this dependence using the notation $w \equiv \beta
  W_{\protect \textrm {diss}}$.}}\BibitemShut {Stop}%
\bibitem [{\citenamefont {Kawai}\ \emph {et~al.}(2007)\citenamefont {Kawai},
  \citenamefont {Parrondo},\ and\ \citenamefont {den Broeck}}]{KPB1}%
  \BibitemOpen
  \bibfield  {author} {\bibinfo {author} {\bibfnamefont {R.}~\bibnamefont
  {Kawai}}, \bibinfo {author} {\bibfnamefont {J.~M.~R.}\ \bibnamefont
  {Parrondo}}, \ and\ \bibinfo {author} {\bibfnamefont {C.~V.}\ \bibnamefont
  {den Broeck}},\ }\href {\doibase 10.1103/PhysRevLett.98.080602} {\bibfield
  {journal} {\bibinfo  {journal} {Phys. Rev. Lett.}\ }\textbf {\bibinfo
  {volume} {98}},\ \bibinfo {pages} {080602} (\bibinfo {year}
  {2007})}\BibitemShut {NoStop}%
\bibitem [{\citenamefont {Yunger~Halpern}\ \emph {et~al.}(2018)\citenamefont
  {Yunger~Halpern}, \citenamefont {Garner}, \citenamefont {Dahlsten},\ and\
  \citenamefont {Vedral}}]{NYH1}%
  \BibitemOpen
  \bibfield  {author} {\bibinfo {author} {\bibfnamefont {N.}~\bibnamefont
  {Yunger~Halpern}}, \bibinfo {author} {\bibfnamefont {A.~J.~P.}\ \bibnamefont
  {Garner}}, \bibinfo {author} {\bibfnamefont {O.~C.~O.}\ \bibnamefont
  {Dahlsten}}, \ and\ \bibinfo {author} {\bibfnamefont {V.}~\bibnamefont
  {Vedral}},\ }\href {\doibase 10.1103/PhysRevE.97.052135} {\bibfield
  {journal} {\bibinfo  {journal} {Phys. Rev. E}\ }\textbf {\bibinfo {volume}
  {97}},\ \bibinfo {pages} {052135} (\bibinfo {year} {2018})}\BibitemShut
  {NoStop}%
\bibitem [{\citenamefont {van Erven}\ and\ \citenamefont
  {Harremos}(2014)}]{NV1}%
  \BibitemOpen
  \bibfield  {author} {\bibinfo {author} {\bibfnamefont {T.}~\bibnamefont {van
  Erven}}\ and\ \bibinfo {author} {\bibfnamefont {P.}~\bibnamefont
  {Harremos}},\ }\href {\doibase 10.1109/TIT.2014.2320500} {\bibfield
  {journal} {\bibinfo  {journal} {IEEE Transactions on Information Theory}\
  }\textbf {\bibinfo {volume} {60}},\ \bibinfo {pages} {3797} (\bibinfo {year}
  {2014})}\BibitemShut {NoStop}%
\bibitem [{\citenamefont {Cover}\ and\ \citenamefont {Thomas}(2005)}]{CT}%
  \BibitemOpen
  \bibfield  {author} {\bibinfo {author} {\bibfnamefont {T.~M.}\ \bibnamefont
  {Cover}}\ and\ \bibinfo {author} {\bibfnamefont {J.~A.}\ \bibnamefont
  {Thomas}},\ }\href {\doibase 10.1002/047174882x} {\emph {\bibinfo {title}
  {Elements of Information Theory}}}\ (\bibinfo  {publisher} {Wiley},\ \bibinfo
  {year} {2005})\BibitemShut {NoStop}%
\bibitem [{\citenamefont {von Neumann}\ and\ \citenamefont
  {Morgenstern}(2007)}]{risk_vNM}%
  \BibitemOpen
  \bibfield  {author} {\bibinfo {author} {\bibfnamefont {J.}~\bibnamefont {von
  Neumann}}\ and\ \bibinfo {author} {\bibfnamefont {O.}~\bibnamefont
  {Morgenstern}},\ }\href {\doibase 10.1515/9781400829460} {\emph {\bibinfo
  {title} {Theory of Games and Economic Behavior (60th Anniversary
  Commemorative Edition)}}}\ (\bibinfo  {publisher} {Princeton University
  Press},\ \bibinfo {year} {2007})\BibitemShut {NoStop}%
\bibitem [{\citenamefont {Bernoulli}(1954)}]{risk_bernoulli}%
  \BibitemOpen
  \bibfield  {author} {\bibinfo {author} {\bibfnamefont {D.}~\bibnamefont
  {Bernoulli}},\ }\href {\doibase 10.2307/1909829} {\bibfield  {journal}
  {\bibinfo  {journal} {Econometrica}\ }\textbf {\bibinfo {volume} {22}},\
  \bibinfo {pages} {23} (\bibinfo {year} {1954})}\BibitemShut {NoStop}%
\bibitem [{\citenamefont {Arrow}(1965)}]{risk_arrow}%
  \BibitemOpen
  \bibfield  {author} {\bibinfo {author} {\bibfnamefont {K.}~\bibnamefont
  {Arrow}},\ }\href {https://books.google.com.co/books?id=hnNEAAAAIAAJ} {\emph
  {\bibinfo {title} {Aspects of the theory of risk-bearing}}}\ (\bibinfo
  {publisher} {Yrj{\"o} Jahnssonin S{\"a}{\"a}ti{\"o}},\ \bibinfo {year}
  {1965})\BibitemShut {NoStop}%
\bibitem [{\citenamefont {Pratt}(1964)}]{risk_pratt}%
  \BibitemOpen
  \bibfield  {author} {\bibinfo {author} {\bibfnamefont {J.~W.}\ \bibnamefont
  {Pratt}},\ }\href {\doibase 10.2307/1913738} {\bibfield  {journal} {\bibinfo
  {journal} {Econometrica}\ }\textbf {\bibinfo {volume} {32}},\ \bibinfo
  {pages} {122} (\bibinfo {year} {1964})}\BibitemShut {NoStop}%
\bibitem [{\citenamefont {de~Finetti}(1952)}]{risk_finetti}%
  \BibitemOpen
  \bibfield  {author} {\bibinfo {author} {\bibfnamefont {B.}~\bibnamefont
  {de~Finetti}},\ }\href {https://www.jstor.org/stable/23236169} {\bibfield
  {journal} {\bibinfo  {journal} {Giornale degli Economisti e Annali di
  Economia}\ }\textbf {\bibinfo {volume} {11}},\ \bibinfo {pages} {685}
  (\bibinfo {year} {1952})}\BibitemShut {NoStop}%
\bibitem [{\citenamefont {{Kelly}}(1956)}]{kelly}%
  \BibitemOpen
  \bibfield  {author} {\bibinfo {author} {\bibfnamefont {J.~L.}\ \bibnamefont
  {{Kelly}}},\ }\href@noop {} {\bibfield  {journal} {\bibinfo  {journal} {The
  Bell System Technical Journal}\ }\textbf {\bibinfo {volume} {35}},\ \bibinfo
  {pages} {917} (\bibinfo {year} {1956})}\BibitemShut {NoStop}%
\bibitem [{\citenamefont {Ducuara}\ and\ \citenamefont
  {Skrzypczyk}(2022)}]{DS2022}%
  \BibitemOpen
  \bibfield  {author} {\bibinfo {author} {\bibfnamefont {A.~F.}\ \bibnamefont
  {Ducuara}}\ and\ \bibinfo {author} {\bibfnamefont {P.}~\bibnamefont
  {Skrzypczyk}},\ }\href {\doibase 10.1103/PRXQuantum.3.020366} {\bibfield
  {journal} {\bibinfo  {journal} {PRX Quantum}\ }\textbf {\bibinfo {volume}
  {3}},\ \bibinfo {pages} {020366} (\bibinfo {year} {2022})}\BibitemShut
  {NoStop}%
\bibitem [{\citenamefont {Soklakov}(2020)}]{risk_soklakov}%
  \BibitemOpen
  \bibfield  {author} {\bibinfo {author} {\bibfnamefont {A.~N.}\ \bibnamefont
  {Soklakov}},\ }\href {\doibase 10.3390/e22080860} {\bibfield  {journal}
  {\bibinfo  {journal} {Entropy}\ }\textbf {\bibinfo {volume} {22}} (\bibinfo
  {year} {2020}),\ 10.3390/e22080860}\BibitemShut {NoStop}%
\bibitem [{\citenamefont {Bleuler}\ \emph {et~al.}(2020)\citenamefont
  {Bleuler}, \citenamefont {Lapidoth},\ and\ \citenamefont {Pfister}}]{BLP1}%
  \BibitemOpen
  \bibfield  {author} {\bibinfo {author} {\bibfnamefont {C.}~\bibnamefont
  {Bleuler}}, \bibinfo {author} {\bibfnamefont {A.}~\bibnamefont {Lapidoth}}, \
  and\ \bibinfo {author} {\bibfnamefont {C.}~\bibnamefont {Pfister}},\ }\href
  {\doibase 10.3390/e22030316} {\bibfield  {journal} {\bibinfo  {journal}
  {Entropy}\ }\textbf {\bibinfo {volume} {22}},\ \bibinfo {pages} {316}
  (\bibinfo {year} {2020})}\BibitemShut {NoStop}%
\bibitem [{Note2()}]{Note2}%
  \BibitemOpen
  \bibinfo {note} {Note that it is more standard to see the ARA defined with an
  additional minus sign. Here, we since we are considering liabilities instead
  of gains, it is natural to omit the minus sign. This means that risk-averse
  agents will have a positive ARA, while risk-seeking agents will have a
  negative ARA, in line with the definitions for gains.}\BibitemShut {Stop}%
\bibitem [{\citenamefont {Wei}\ and\ \citenamefont
  {Plenio}(2017)}]{Wei_Plenio}%
  \BibitemOpen
  \bibfield  {author} {\bibinfo {author} {\bibfnamefont {B.-B.}\ \bibnamefont
  {Wei}}\ and\ \bibinfo {author} {\bibfnamefont {M.~B.}\ \bibnamefont
  {Plenio}},\ }\href {\doibase 10.1088/1367-2630/aa579e} {\bibfield  {journal}
  {\bibinfo  {journal} {New Journal of Physics}\ }\textbf {\bibinfo {volume}
  {19}},\ \bibinfo {pages} {023002} (\bibinfo {year} {2017})}\BibitemShut
  {NoStop}%
\bibitem [{\citenamefont {Buscemi}\ and\ \citenamefont
  {Scarani}(2021)}]{BS2021}%
  \BibitemOpen
  \bibfield  {author} {\bibinfo {author} {\bibfnamefont {F.}~\bibnamefont
  {Buscemi}}\ and\ \bibinfo {author} {\bibfnamefont {V.}~\bibnamefont
  {Scarani}},\ }\href {\doibase 10.1103/PhysRevE.103.052111} {\bibfield
  {journal} {\bibinfo  {journal} {Phys. Rev. E}\ }\textbf {\bibinfo {volume}
  {103}},\ \bibinfo {pages} {052111} (\bibinfo {year} {2021})}\BibitemShut
  {NoStop}%
\bibitem [{\citenamefont {Aw}\ \emph {et~al.}(2021)\citenamefont {Aw},
  \citenamefont {Buscemi},\ and\ \citenamefont {Scarani}}]{ABS}%
  \BibitemOpen
  \bibfield  {author} {\bibinfo {author} {\bibfnamefont {C.~C.}\ \bibnamefont
  {Aw}}, \bibinfo {author} {\bibfnamefont {F.}~\bibnamefont {Buscemi}}, \ and\
  \bibinfo {author} {\bibfnamefont {V.}~\bibnamefont {Scarani}},\ }\href@noop
  {} {\bibfield  {journal} {\bibinfo  {journal} {AVS Quantum Sci.}\ }\textbf
  {\bibinfo {volume} {3}},\ \bibinfo {pages} {045601} (\bibinfo {year}
  {2021})}\BibitemShut {NoStop}%
\bibitem [{\citenamefont {Sason}\ and\ \citenamefont {Verdú}(2016)}]{FD}%
  \BibitemOpen
  \bibfield  {author} {\bibinfo {author} {\bibfnamefont {I.}~\bibnamefont
  {Sason}}\ and\ \bibinfo {author} {\bibfnamefont {S.}~\bibnamefont {Verdú}},\
  }\href {\doibase 10.1109/TIT.2016.2603151} {\bibfield  {journal} {\bibinfo
  {journal} {IEEE Transactions on Information Theory}\ }\textbf {\bibinfo
  {volume} {62}},\ \bibinfo {pages} {5973} (\bibinfo {year}
  {2016})}\BibitemShut {NoStop}%
\bibitem [{\citenamefont {Sagawa}(2013)}]{sagawa2013}%
  \BibitemOpen
  \bibfield  {author} {\bibinfo {author} {\bibfnamefont {T.}~\bibnamefont
  {Sagawa}},\ }\href {\doibase 10.1007/978-4-431-54168-4} {\emph {\bibinfo
  {title} {Thermodynamics of Information Processing in Small Systems}}}\
  (\bibinfo  {publisher} {Springer Japan},\ \bibinfo {year} {2013})\BibitemShut
  {NoStop}%
\bibitem [{\citenamefont {Morales}\ \emph {et~al.}(2023)\citenamefont
  {Morales}, \citenamefont {Korbel},\ and\ \citenamefont {Rosas}}]{morales23}%
  \BibitemOpen
  \bibfield  {author} {\bibinfo {author} {\bibfnamefont {P.~A.}\ \bibnamefont
  {Morales}}, \bibinfo {author} {\bibfnamefont {J.}~\bibnamefont {Korbel}}, \
  and\ \bibinfo {author} {\bibfnamefont {F.~E.}\ \bibnamefont {Rosas}},\ }\href
  {\doibase 10.1088/1367-2630/ace4eb} {\bibfield  {journal} {\bibinfo
  {journal} {New Journal of Physics}\ }\textbf {\bibinfo {volume} {25}},\
  \bibinfo {pages} {073011} (\bibinfo {year} {2023})}\BibitemShut {NoStop}%
\bibitem [{\citenamefont {Caravelli}(2022)}]{caravelli22}%
  \BibitemOpen
  \bibfield  {author} {\bibinfo {author} {\bibfnamefont {F.}~\bibnamefont
  {Caravelli}},\ }\href@noop {} {\enquote {\bibinfo {title} {The bellman
  equation and optimal local flipping strategies for kinetic ising models},}\ }
  (\bibinfo {year} {2022}),\ \Eprint {http://arxiv.org/abs/arXiv:2201.02081}
  {arXiv:2201.02081} \BibitemShut {NoStop}%
\bibitem [{\citenamefont {Campbell}(1965)}]{campbell65}%
  \BibitemOpen
  \bibfield  {author} {\bibinfo {author} {\bibfnamefont {L.}~\bibnamefont
  {Campbell}},\ }\href {\doibase 10.1016/s0019-9958(65)90332-3} {\bibfield
  {journal} {\bibinfo  {journal} {Information and Control}\ }\textbf {\bibinfo
  {volume} {8}},\ \bibinfo {pages} {423} (\bibinfo {year} {1965})}\BibitemShut
  {NoStop}%
\end{thebibliography}%

\end{document}